\documentclass{elsarticle}
\usepackage{latexsym}
\usepackage{float}
\usepackage[latin2]{inputenc}
\usepackage{amssymb}
\usepackage{amsthm}
\newtheorem{theorem}{Theorem}

\newtheorem{definition}{Definition}
\newtheorem{corollary}[theorem]{Corollary}
\newtheorem{conjecture}[theorem]{Conjecture}

\newtheorem{example}[theorem]{Example}

\newtheorem{remark}[theorem]{Remark}

\def\rank{\hbox{\rm rank}}

\usepackage[hidelinks]{hyperref}

\makeatletter
\def\ps@pprintTitle{%
	\let\@oddhead\@empty
	\let\@evenhead\@empty
	\def\@oddfoot{\centerline{\thepage}}%
	\let\@evenfoot\@oddfoot}
\renewcommand\@biblabel[1]{(#1)} 
\makeatother
\makeatletter
\gdef\emailauthor#1#2{\stepcounter{ead}%
     \g@addto@macro\@elseads{\raggedright%
      \let\corref\@gobble
      \eadsep\texttt{#1}\def\eadsep{\unskip,\space}}%
}
\def\urlauthor#1#2{\g@addto@macro\@elsuads{\let\corref\@gobble%
    \raggedright\eadsep\texttt{#1}%
    \def\eadsep{\unskip,\space}}%
}
\makeatother

\begin{document}

\title{A Note on the LogRank Conjecture in Communication Complexity}

\author[a]{Vince Grolmusz
\ead{grolmusz@pitgroup.org}
\address[a]{E\"otv\"os University, H-1117 Budapest, Hungary}}

\date{}

\begin{abstract}
The LogRank conjecture of Lovász and Saks from 1988 is the most famous open problem in the communication complexity theory. The statement is as follows: Suppose that two players intend to compute a Boolean function $f(x,y)$ when $x$ is known for the first and $y$ for the second player, and they may send and receive messages encoded with bits, then they can compute $f(x,y)$ with exchanging $(\log \rank (M_f))^c $ bits, where $M_f$ is a Boolean matrix, determined by function $f$. The problem is widely open and very popular, and it has resisted numerous attacks in the last 35 years. The best upper bound is still exponential in the bound of the conjecture. Unfortunately, we cannot prove the conjecture, but we present a communication protocol with $(\log \rank (M_f))^c $ bits, which computes a -- somewhat -- related quantity to $f(x,y)$. The relation is characterized by a representation of low-degree, multi-linear polynomials modulo composite numbers. This result of ours may help to settle this long-time open conjecture.
\end{abstract}

\maketitle
\section{Introduction}

\subsection{Communication Games}

Two-player communication games were first defined by Yao in 1979 \cite{Y79}: There are two players, Alice and Bob, and a Boolean function $$f:\{0,1\}^N\times\{0,1\}^N\to\{0,1\},$$
and two Boolean sequences, $x,y\in\{0,1\}^N$. Alice knows the value of $x$, Bob the value of $y$, and they want to compute collaboratively the value of $f(x,y)$. The local computational power of the players are unlimited, and the cost of the collaborative computation is the number of bits exchanged between the parties. Function $f$ is computed by the players if one of them knows $f(x,y)$, and the other knows that the first player knows \cite{Lov-sur}. 

Clearly, any $f$ can be computed by $N$ bits of communication: Alice tells $x$ to Bob at cost $N$, and Bob computes $f(x,y)$ for free. This is the trivial protocol for computing $f$.

We say that the players follow a {\it communication protocol} for computing $f$, where the protocol prescribes that in each step, what message a player should send to the other for a given message history and input. The cost of a protocol, computing $f(x,y)$, is the maximum number of bits communicated, taken for all $x$ and $y$ inputs. 

The communication complexity of Boolean function $f$ is the minimum cost of protocols, which compute $f$. The communication complexity of $f$ is denoted by $\kappa(f)$ \cite{Lov-sur}. For a more formal introduction and examples, we refer to \cite{Lov-sur,NiKu}.

The communication games and communication complexity have become a central field of theoretical computer science, hundreds of publications (e.g., \cite{N,NW,BNS,G94a, Lovett2016}  and numerous books \cite{NiKu,Rao2020,Karchmer1989} have been appeared on the topic. For example, difficult-to-handle areas, such as Boolean circuit complexity, applied communication games for gaining upper- and lower bounds \cite{HG,G98,G-dede,G92,G95a,GrT,G-HARM,Karchmer1989,knsw}.

One of the main questions is finding general upper- and lower bounds for the communication complexity of $f$. To describe the bounds, we need to define the communication matrix of the function $f$:

\begin{definition}
The communication matrix of $f:\{0,1\}^N\times\{0,1\}^N\to\{0,1\}$ is a $2^N\times2^N$ 0-1 matrix $M_f$, where its rows corresponds to the different $x\in \{0,1\}^N$ values, the columns to the different $y\in \{0,1\}^N$ values, and in the position of row $x$ and column $y$ is the value of $f(x,y)\in\{0,1\}$. Let $\rank(M_f)$ denote the matrix rank over the rational field. Let $\log$ denote the logarithm base 2, and $\ln $ the natural logarithm.
\end{definition}

A general lower bound, which implies that for most of the natural communications problems (e.g., the identity function or the set disjointness) the trivial protocol is optimal, was proved by Mehlhorn and Schmidt in 1982:

\begin{theorem}[\cite{MS}] 
If $f$ is not the identically $0$ function, then $$\kappa(f)\geq \log\rank(M_f).$$
\end{theorem}

The most famous open problem in communication complexity is the LogRank conjecture of Lovász and Saks from 1988:

\begin{conjecture}[\cite{LS}]\label{conj}
 There exists a polynomial $P$ such that for all $f\not \equiv 0$, 
$$\kappa(f)\leq P(\log \rank (M_f)).$$
\end{conjecture}

The 35-year-old conjecture is widely open today, inspiring numerous theorems and approaches even in the last few years \cite{Singer2022,Wu2021,Gal2021,Knop2021,Chattopadhyay2020,Kol2018,Sinha2019}. 

The best published upper bound of Lovett \cite{Lovett2016} (for a large enough rank($M_f$)) is still exponentially far from the conjectured upper bound:
$$\kappa(f)=O\left(\sqrt{\rank(M_f)}\log\rank(M_f)\right).$$

We note that the LogRank conjecture can be formulated with the terms of graph theory as a relation between the rank of the adjacency matrix and the chromatic number of a graph \cite{LS,Lovett2016}, without even mentioning communication games.

\subsection{Polynomial representations modulo composite numbers}

We will need some definitions and theorems from \cite{G-six,Grolmusz2008}:

Let $g:\{0,1\}^n\to\{0,1\}$ be a Boolean function and let $m$ be a
positive integer. {\em Barrington}, {\em Beigel} and {\em Rudich}
\cite{BBR} gave the following definition:

\begin{definition}\label{weekly}
The polynomial $P$ with integer coefficients {\em weakly represents}
 Boolean function $g$ modulo $m$ if there exists an
$S\subset \{0,1,2,...,m-1\}$ such that for all $x\in\{0,1\}^n$, 
$$g(x)=0\Longleftrightarrow \big(P(x)\bmod{m}\big)\in S.$$
Here $(a \bmod{m})$ denotes the smallest non-negative $b\equiv a\bmod{m}$.
\end{definition}

We are interested in the smallest degree polynomials representing $g$. Since $g$ is Boolean, we may assume that $P$ is multilinear (since $x_i^2=x_i$ over $\{0,1\}^n$).

\noindent Let $\hbox{OR}_n:\{0,1\}^n\to\{0,1\}$ denote the $n$--variable OR-function:
$$\hbox{OR}_n(x_1,x_2,\ldots,x_n)=\cases{0, \hbox{ if }x_1=x_2=\cdots=x_n=0\cr
                                         1\hbox{ otherwise.}}$$

If polynomial $P$ weakly represents OR$_n$ modulo a prime $p$ then we may assume that for 
$x\in\{0,1\}^n$,

$$P(x)= 0\bmod{p}\iff x=(0,0,...,0).$$ 

Then 
$$1-P^{p-1}(1-x_1,1-x_2,...,1-x_n)$$
is clearly  the $n$-variable AND function with a unique multi-linear form
$$x_1x_2x_3...x_n.$$ 
Therefore, the degree of $P$ is at least $n/(p-1)$.

{\em Tardos} and {\em Barrington} \cite{TB} showed that the same conclusion holds if $p$ is a prime power.

However, {\em Barrington, Beigel} and {\em Rudich} \cite{BBR} proved that the conclusion fails for composite
moduli with at least two prime divisors:

\begin{theorem}[\cite{BBR}]\label{egy}
There exists an explicitly constructible polynomial $P$ of degree
$O(n^{1/r})$ which weakly represents $OR_n$ modulo $m=p_1^{\alpha_1}p_2^{\alpha_2}...p_r^{\alpha_r}$, 
where the $p_i$'s are distinct primes.
\end{theorem}

An explicit example of such a non-trivial polynomial mod 6 is given in the Appendix.

We have applied the polynomial of Theorem \ref{egy} in \cite{G99} for falsifying a long-standing conjecture for the size of set systems with restricted intersections and also for giving new explicit Ramsey-graph constructions, among other applications for set systems and codes described in \cite{G-hyram, G-polyset, Grolmusz2003, Grolmusz2006,DMTCS,Grolmusz2006a}.

In \cite{G99} we have reproduced a short proof of Theorem \ref{egy} of \cite{BBR}, and we have proved the following 

\begin{corollary}\label{coro}
Let $m=p_1^{\alpha_1}p_2^{\alpha_2}...p_r^{\alpha_r}$. Then there
exists an explicitly constructible polynomial $P'$ with $n$ variables
and of degree $O(n^{1/r})$ which is equal to 0 on $x=(0,0,\ldots,0)
\in\{0,1\}^n$, it is nonzero mod~$m$ for all other $x\in\{0,1\}^n$,
and for all $x\in\{0,1\}^n$ and for all $i\in\{1,\ldots,r\}$,
$P(x)\equiv0\pmod{p_i^{\alpha_i}}$ or
  $P(x)\equiv1\pmod{p_i^{\alpha_i}}$.
\end{corollary}

Using the results of \cite{G99}, we have found a remarkable application for elementary symmetric polynomials in \cite{G-six}. 

We will need the following definition from \cite{Grolmusz2008} with small changes to describe the result:

\begin{definition}\label{a-strong}
Let $m$ be a composite number with prime-factorization $m=p_1^{e_1}p_2^{e_2}\cdots 
p_\ell^{e_\ell}$. Let $\mathbb{Z}_m$ denote the ring of integers modulo $m$ and $\mathbb{Z}$ the integers. Let 
$f$ be a multi-linear polynomial of $n$ variables over the integers:
$$f(x_1,x_2,\ldots,x_n)=\sum_{\alpha\in \{0,1\}^n}a_\alpha x^\alpha,$$
where $a_\alpha\in \mathbb{Z}$, $x^\alpha=\prod_{i=1}^n x_i^{\alpha_i}$. Then we say that
$$g(x_1,x_2,\ldots,x_n)=\sum_{\alpha\in \{0,1\}^n}b_\alpha x^\alpha,$$
is an
\begin{itemize}
\item {\em alternative representation} of $f$ modulo $m$, if
$$\forall \alpha\in \{0,1\}^n \ \  \exists j\in \{1,2,\ldots,\ell\}:$$ 
$$ a_\alpha\equiv b_\alpha \pmod{p_j^{e_j}};$$
\item {\em 0-a-strong representation} of $f$ modulo $m$, if
it is an alternative representation, and furthermore,
 if for some $i$, $a_\alpha\not\equiv b_\alpha \pmod{p_i^{e_i}},$ then 
$b_\alpha\equiv 0\pmod{p_i^{e_i}};$
\item {\em 1-a-strong representation} of $f$ modulo $m$, if
it is an alternative representation, and furthermore,
 if for some $i$, $a_\alpha\not\equiv b_\alpha \pmod{p_i^{e_i}},$ then 
$a_\alpha\equiv 0\pmod{m};$

\end{itemize}
\end{definition}

That is for modulus 6; in the alternative representation, each coefficient is correct, either modulo 2 or modulo 3, but not necessarily both. 

In the 0-a-strong representation, the 0 coefficients are always correct both modulo 2 and 3; the non-zeroes are allowed to be correct either modulo 2 or 3, and if they are not correct modulo one of them, say 2, then they should be 0 mod 2. Consequently, the coefficient 1 can be represented by 1, 3, or 4, but nothing else.

In the 1-a-strong representation, the non-zero coefficients of $f$ are correct for both moduli in $g$, but the zero coefficients of $f$ can be non-zero either modulo 2 or modulo 3 in $g$, but not both.

\begin{remark}[\cite{Grolmusz2008}] The 1-a-strong representations of polynomial $f$ can be written in the form modulo $m$:
$$f+p_1^{e_1}g_1+p_2^{e_2}g_2+\cdots+p_\ell^{e_\ell}g_\ell,$$
where the $g_i$ have no monomials in common with each other, nor with $f$.
\end{remark}

\begin{example}[\cite{Grolmusz2008}]Let $m=6$, and let $f(x_1,x_2,x_3)=x_1x_2+x_2x_3+x_1x_3$,
then 
$g(x_1,x_2,,x_3)=3x_1x_2+4x_2x_3+x_1x_3$ is a 0-a-strong representation 
of $f$ modulo 6;
$g(x_1,x_2,,x_3)=x_1x_2+x_2x_3+x_1x_3+3x_1^2+4x_2$
is a 1-a-strong representation of $f$ modulo 6;
$g(x_1,x_2,,x_3)=3x_1x_2+4x_2x_3+x_1x_3+3x_1^2+4x_2$
is an alternative representation modulo 6.
\end{example}

\begin{remark} Clearly, the 1-a-strong representation of $f$ is not unique. Suppose that 
\begin{itemize}
\item[(i)] all coefficients of $f$ and $f'$ are either 1 or -1 mod $m$, and

\item[(ii)] $g$ is a 1-a-strong representation of $f$ and also of $f'$, where $f,f'$ and $g$ are multilinear, homogeneous degree-$d$ polynomials, that is, every monomials of $f,f'$ and $g$ are degree $d$, 
\end{itemize}

then $f=f'$ modulo $m$. Clearly, one can set the monomials of $f$ or $f'$ to 1 one by one, and the $p_i$-multiplied monomials need to be 0 in $g$ because of homogeneity. 
\end{remark}

We have proved in \cite{G-six} and stated in this form in \cite{Grolmusz2008} the following

\begin{theorem}[\cite{G-six}]\label{fo-11}
 Let the prime factorization of positive integer $m$ be $m=p_1^{e_1}p_2^{e_2}\cdots p_\ell^{e_\ell}$, where $\ell>1$. Then a degree-2 0-a-strong representation of the second elementary symmetric polynomial
$$S_n^2(x,y)=\sum_{i,j\in\{1,2,\ldots,n\}\atop{i\ne j}}x_iy_j,\eqno{(1)}$$
 modulo $m$:
$$\sum_{i,j\in\{1,2,\ldots,n\}\atop{i\ne j}}a_{ij}x_iy_j\eqno{(2)}$$
 can be computed as the following product with  coefficients from $\mathbb{Z}_m$:
$$\sum_{j=1}^{t}\left(\sum_{i=1}^nb'_{ij}x_i\right)\left(\sum_{i=1}^nc'_{ij}y_i\right)$$
where  $t=\exp(O(\sqrt[\ell]{\log n(\log \log n)^{\ell-1}})).$
Moreover, this representation satisfies that $\forall i\ne j: a_{ij}=a_{ji}$.
\end{theorem}\qed

Now we need the main theorem from \cite{Grolmusz2008}:

\begin{theorem}[\cite{Grolmusz2008}, Theorem 6]\label{dot}
Let $m=p_1^{e_1}p_2^{e_2}\cdots p_\ell^{e_\ell}$, where $\ell>1$, and $p_1,p_2,\ldots,p_\ell$ are primes. Then a degree-2, 1-a-strong representation of the dot-product
$f(x_1,x_2,\ldots,x_n,y_1,y_2,\ldots,y_n)=\sum_{i=1}^n x_iy_i$
can be computed with 
$t+1=\exp(O(\sqrt[\ell]{\log n(\log \log n)^{\ell-1}}))$
multiplications of the form 
$$\sum_{j=1}^{t+1}\left(\sum_{i=1}^nb_{ij}x_i\right)\left(\sum_{i=1}^nc_{ij}y_i\right),\eqno{(3)}$$
where all coefficients are integers.
\end{theorem}

The proof is immediate by subtracting the 0-a-strong representation of $S_n^2(x,y)$ from $(x_1+x_2+\ldots+x_n)(y_1+y_2+\ldots+y_n)$ \cite{Grolmusz2008}.

An explicit example for the $b_{ij}$ and $c_{ij}$ coefficients, with $m=6,n=16,t+1=13$ can be found in the Appendix.

\section{The LogRank Protocol}

Here, we describe our protocol. Let $m=p_1p_2\ldots p_\ell$ be the product of the first $\ell$ primes. Then $m=e^{(1+o(1))\ell\ln \ell}$, by the estimation of the first Chebyshev number \cite{Hardy1995}. Let $\ell=\lfloor  \log \log n\rfloor $. Then $m$ can be given with less than $(\log \log n)^c$ bits, with a $c>0$.

The value $t$ from Theorem \ref{dot} satisfies 
$$t\leq  \exp ( O((\log n)^{1/\ell}\log \log n))\leq \exp (c_1 \log \log n) \leq (\log n)^{c_1},$$
with a positive $c_1$.

Now suppose that we have a Boolean function $F:\{0,1\}^N\times \{0,1\}^N\to \{0,1\}$, with a 0-1 $2^N\times 2^N$ communication matrix $M_F$ with $\rank(M_F)=n$ over the rationals, $n\leq 2^N$. By the definition of the rank, one may choose $n$ linearly independent 0-1 columns of $M_F$, let the $2^N\times n$ 0-1 matrix $X$ contain these columns. Then there exists an $n\times 2^N$ rational matrix $Y$, such that $M_F=XY$.

Note that each entry of $M_F$ is a dot product of a length-$n$ row of $X$ and a length-$n$ column of $Y$. Now, by Theorem \ref{dot}, the 1-a-strong representation of this dot product can be computed modulo $m$ as a sum (3):

$$\sum_{j=1}^{t+1}\left(\sum_{i=1}^nb_{ij}x_i\right)\left(\sum_{i=1}^nc_{ij}y_i\right),$$

where all coefficients are integers modulo $m$. Now, Alice can compute privately for $j=1,2,\ldots t+1$ the mod $m$ values of sums
$$\left(\sum_{i=1}^nb_{ij}x_i\right)$$
and can communicate each with $(\log \log n)^c$ bits. Since $t<(\log n)^{c_1}$, the total communication of Alice is polylogarithmic in the rank $n$. 

Now, Bob, knowing the (rational) values of $y_1,y_2,\ldots,y_n$ can privately compute the sum (3), without further communication.

\begin{remark}
	For a given $M_f$, the players can agree on that the will compute the $kF(x,y)$ instead of $F(x,y)$, and then matrix $Y$ can be changed for an integer matrix $kY$. Note also that this convention will not increase the communication, since Bob does not communicate linear combinations of his variables.	
\end{remark}

\section{Remarks and Conclusion}

In the previous section we have presented a protocol, which computed a value with $(\log \rank(M_F))^{c_2}$ communication, with a positive constant $c_2$ which is substituted in the 1-a-strong representation $g$ modulo $m=p_1p_2\ldots p_\ell$ of the dot-product polynomial $f=\sum_{i=1}^n x_iy_i$:
$$g(x,y)=\sum_{i=1}^n x_iy_i+p_1g_1(x,y)+p_2g_2(x,y)+\ldots+p_\ell g_\ell(x,y).\eqno{(4)}$$

Note that the "surplus" terms in (4) are zero modulo for at least one of the prime divisors of $m$. Note also that there is no common monomial (i.e., $x_iy_j$) in the dot-product and among the $g_i$ polynomials. 

\begin{remark}
We do not know how to eliminate the surplus terms with the $g_i$ polynomials from $(4)$ with a log-rank bounded additional communication. One can imagine numerous possible approaches, for example, substituting a prime $p_i, 1\leq i\leq \ell$ for all or just a subset of 1's in the $x_i$'s, and repeating the protocol; or repeating the protocol above for 0-1 rows instead of 0-1 columns, or repeating the protocol above for overlaying subsets of indices $i=1,2,\ldots,n$ several times.
\end{remark}

\begin{remark}
As we described in \cite{Grolmusz2008}, Theorem \ref{dot} has numerous applications in representing the matrix product with very few 
multiplications or the hyperdense coding of numbers, vectors or matrices. We also note that our definition and use of the term "hyperdense coding" \cite{Grolmusz2003a} precede the quantum-computational use of an unrelated but identically named term of \cite{Imre2012,Bebrov19} by more than 9 years.
\end{remark}

\section*{Funding}

VG was partially funded by the Ministry of Innovation and Technology of Hungary from the National Research, Development and Innovation Fund, financed under the  ELTE TKP 2021-NKTA-62 funding scheme.

\section*{Competing interests} The author declares no competing interests.
%\section*{References}

%\bibliography{v:/vince/CIKKEK/medl}
%\bibliographystyle{unsrtnat}

\section{Appendix}

Here we give an example for a polynomial what we have presented in \cite{G99}:

\begin{example}[\cite{G99}]
 Let $m=6$, and let 
$$G_1(x)=\sum^{2^3-1}_{j=1}(-1)^{j+1}s_j(x),$$
and
$$G_2(x)=\sum^{3^2-1}_{j=1}(-1)^{j+1}s_j(x).$$
Then
$$P(x)=3G_1(x)+4G_2(x)$$
weakly represents  $OR_{71}$  modulo 6, and its degree is only 8. 
\end{example}

\begin{example}

Here we give an explicit example for the coefficients in Theorem \ref{dot} (3).

Let $m=6, n=16, t+1=13$ and let $B$ be a $16\times 13$ matrix, $C$ be a $13\times 16$ matrix mod 6, then $A=BC$ mod 6 is a $16\times 16$ matrix with 1's in the main diagonal and 0's either mod 2 or mod 3 or both outside the main diagonal:

$$B=\left[ \begin {array}{ccccccccccccc} 1&0&1&1&4&1&4&4&3&1&4&4&4
\\ \noalign{\medskip}1&1&0&4&1&4&1&3&4&4&1&3&3\\ \noalign{\medskip}1&1
&4&0&1&4&3&1&4&4&3&1&3\\ \noalign{\medskip}1&4&1&1&0&3&4&4&1&3&4&4&4
\\ \noalign{\medskip}1&1&4&4&3&0&1&1&4&4&3&3&1\\ \noalign{\medskip}1&4
&1&3&4&1&0&4&1&3&4&4&4\\ \noalign{\medskip}1&4&3&1&4&1&4&0&1&3&4&4&4
\\ \noalign{\medskip}1&3&4&4&1&4&1&1&0&4&3&3&3\\ \noalign{\medskip}1&1
&4&4&3&4&3&3&4&0&1&1&1\\ \noalign{\medskip}1&4&1&3&4&3&4&4&3&1&0&4&4
\\ \noalign{\medskip}1&4&3&1&4&3&4&4&3&1&4&0&4\\ \noalign{\medskip}1&3
&4&4&1&4&3&3&4&4&1&1&3\\ \noalign{\medskip}1&4&3&3&4&1&4&4&3&1&4&4&0
\\ \noalign{\medskip}1&3&4&4&3&4&1&3&4&4&1&3&1\\ \noalign{\medskip}1&3
&4&4&3&4&3&1&4&4&3&1&1\\ \noalign{\medskip}1&4&3&3&4&3&4&4&1&3&4&4&4
\end {array} \right] $$

$$C=\left[ \begin {array}{cccccccccccccccc} 1&1&1&1&1&1&1&1&1&1&1&1&1&1&1
&1\\ \noalign{\medskip}-1&0&0&0&0&0&0&0&0&0&0&0&0&0&0&-1
\\ \noalign{\medskip}0&-1&0&0&0&0&0&0&0&0&0&0&0&0&-1&0
\\ \noalign{\medskip}0&0&-1&0&0&0&0&0&0&0&0&0&0&-1&0&0
\\ \noalign{\medskip}0&0&0&-1&0&0&0&0&0&0&0&0&0&1&1&1
\\ \noalign{\medskip}0&0&0&0&-1&0&0&0&0&0&0&-1&0&0&0&0
\\ \noalign{\medskip}0&0&0&0&0&-1&0&0&0&0&0&1&0&0&1&1
\\ \noalign{\medskip}0&0&0&0&0&0&-1&0&0&0&0&1&0&1&0&1
\\ \noalign{\medskip}0&0&0&0&0&0&0&-1&0&0&0&-1&0&-1&-1&-2
\\ \noalign{\medskip}0&0&0&0&0&0&0&0&-1&0&0&1&0&1&1&2
\\ \noalign{\medskip}0&0&0&0&0&0&0&0&0&-1&0&-1&0&-1&0&-1
\\ \noalign{\medskip}0&0&0&0&0&0&0&0&0&0&-1&-1&0&0&-1&-1
\\ \noalign{\medskip}0&0&0&0&0&0&0&0&0&0&0&0&-1&-1&-1&-1\end {array}
 \right] $$

$$A=BC=\left[ \begin {array}{cccccccccccccccc} 1&0&0&-3&0&-3&-3&-2&0&-3&-3&-
2&-3&-2&-2&-3\\ \noalign{\medskip}0&1&-3&0&-3&0&-2&-3&-3&0&-2&-3&-2&-3
&-3&-2\\ \noalign{\medskip}0&-3&1&0&-3&-2&0&-3&-3&-2&0&-3&-2&-3&-3&-2
\\ \noalign{\medskip}-3&0&0&1&-2&-3&-3&0&-2&-3&-3&0&-3&-2&-2&-3
\\ \noalign{\medskip}0&-3&-3&-2&1&0&0&-3&-3&-2&-2&-3&0&-3&-3&-2
\\ \noalign{\medskip}-3&0&-2&-3&0&1&-3&0&-2&-3&-3&-2&-3&0&-2&-3
\\ \noalign{\medskip}-3&-2&0&-3&0&-3&1&0&-2&-3&-3&-2&-3&-2&0&-3
\\ \noalign{\medskip}-2&-3&-3&0&-3&0&0&1&-3&-2&-2&-3&-2&-3&-3&0
\\ \noalign{\medskip}0&-3&-3&-2&-3&-2&-2&-3&1&0&0&-3&0&-3&-3&-2
\\ \noalign{\medskip}-3&0&-2&-3&-2&-3&-3&-2&0&1&-3&0&-3&0&-2&-3
\\ \noalign{\medskip}-3&-2&0&-3&-2&-3&-3&-2&0&-3&1&0&-3&-2&0&-3
\\ \noalign{\medskip}-2&-3&-3&0&-3&-2&-2&-3&-3&0&0&1&-2&-3&-3&0
\\ \noalign{\medskip}-3&-2&-2&-3&0&-3&-3&-2&0&-3&-3&-2&1&0&0&-3
\\ \noalign{\medskip}-2&-3&-3&-2&-3&0&-2&-3&-3&0&-2&-3&0&1&-3&0
\\ \noalign{\medskip}-2&-3&-3&-2&-3&-2&0&-3&-3&-2&0&-3&0&-3&1&0
\\ \noalign{\medskip}-3&-2&-2&-3&-2&-3&-3&0&-2&-3&-3&0&-3&0&0&1
\end {array} \right] $$

If we take the dot product of column $j$ of $B$ each with $(x_1,x_2,\ldots,x_n)$ then we will get the $t+1$ left sums of (3):
$$\left(\sum_{i=1}^nb_{ij}x_i\right)$$ 

If we take the dot product of row $j$ of $C$ each with $(x_1,x_2,\ldots,x_n)$ then we will get the $t+1$ right sums of (3):
$$\left(\sum_{i=1}^nc_{ij}y_i\right)$$

Now, clearly, in (3) one can get coefficient of $x_uy_v$ by computing the dot product of row $u$ of $B$ with column $v$ of $C$; that is, matrix $A$ describes the coefficients of the mod 6 representation of the dot product 

$f(x_1,x_2,\ldots,x_n,y_1,y_2,\ldots,y_n)=\sum_{i=1}^n x_iy_i$: all the elements of the main diagonal are 1, and the others are 0 mod 2 or mod 3 or both.

For readers wishing to experiment with these matrices, a Maple worksheet can be accessed at \url{https://grolmusz.pitgroup.org/?attachment_id=1840} 

\end{example}
\end{document}